# Multifrequency Nanoscale Impedance Microscopy (m-NIM): A novel approach towards detection of selective and subtle modifications on the surface of polycrystalline boron-doped diamond electrodes

Artur Zielinski[1], Mateusz Cieslik[1], Michal Sobaszek[2], Robert Bogdanowicz[2], Kazimierz Darowicki[1] and Jacek Ryl*[,1]

[1] Department of Electrochemistry, Corrosion and Materials Engineering, Faculty of Chemistry, Gdansk University of Technology, Narutowicza 11/12, 80-233 Gdansk, Poland

[2] Department of Metrology and Optoelectronics, Faculty of Electronics, Telecommunication and Informatics, Gdansk University of Technology, Narutowicza 11/12, 80-233 Gdansk, Poland

*corresponding author: jacek.ryl@pg.edu.pl

**Abstract**

In this paper, we describe the modification of Nanoscale Impedance Microscopy (NIM), namely, a combination of contact-mode atomic force microscopy with local impedance measurements. The postulated approach is based on the application of multifrequency voltage perturbation instead of standard frequency-by-frequency analysis, which among others offers more time-efficient and accurate determination of the resultant impedance spectra with high spatial resolution. Based on the impedance spectra analysis with an appropriate electric equivalent circuit, it was possible to map surface resistance and contact capacitance.

Polycrystalline heavy boron-doped diamond (BDD) electrodes were the research object. Recent studies have shown that the exposure of such electrodes to oxidizing environment may result in the modification of termination type, and thus it is a key factor in describing the electric and electrochemical properties of BDD. We have successfully applied multifrequency NIM, which allowed us to prove that the modification of termination type is selective and occurs with different propensity on the grains having specific crystallographic orientation. Furthermore, our approach enabled the detection of even subtle submicroscopic surface heterogeneities, created as a result of various oxidation treatments and to distinguish them from the surface heterogeneity related to the local distribution of boron at the grain boundaries.

# 1. Introduction

Atomic force microscopy (AFM) [1] allows the acquisition of sample topography by means of local interactions between the analyzed surface and a sharp probe tip. Such design of the measurement procedure turned out to be fruitful in many fields, ranging from quantum mechanics [2] through classical material engineering [3] and nanomachinery [4] to life sciences [5]. Undoubtedly, one of important advantages of AFM is the possibility of recording various additional interactions (electrostatic [6] or magnetic [7] in origin, for example) between the probe and the sample, synchronously with the topographic image. Methodology utilized in this report is derived from the concept of one of extended contact-mode AFM techniques which is scanning spreading resistance microscopy (SSRM) [8]. In SSRM, the DC voltage is applied between the investigated material and a conductive probe. The mechanical feedback is used to maintain a constant force between the sample and the probe, thus the observed changes in resulting current can be attributed to the heterogeneity of material electrical conductance. The important extension of the discussed electrical measurement methodology was the introduction of AC perturbation, which is referred to as nanoscale impedance microscopy (NIM) [9]. It enabled not only to investigate dielectric materials [10], but also to put into use possibilities offered by impedance spectroscopy [11,12].

The aforementioned technique has found practical research applications, as confirmed by a number of literature reports. Layson et al. presented its application to polymer film research [13]. The work of Pingree et al. [14] focused on technologically important issue of the diagnostics of semiconductor devices, while Kruempelman presented the results on conductive glass [15]. The material heterogeneities were also the subject of numerous studies, i.e. on the grain boundaries [16,17] or local polymer defects [18]. Performance of zinc-rich protective coatings in high-humidity conditions was evaluated using the AFM-based approach [19].

It is our intention to present an innovative approach to combined AFM-impedance measurements, which is based on instantaneous multifrequency impedance measurements, introduced initially in electrochemistry as Dynamic Electrochemical Impedance Spectroscopy (DEIS). The theoretical background of DEIS was first proposed by Darowicki [20] and used in various electrochemical studies of non-stationary systems, in particular to describe corrosion phenomena [21–28]. The application of dynamic impedance spectroscopy to atomic force microscopy was introduced shortly after by Zielinski et al. [29,30], revealing its high application potential for the determination of local disturbances in electric parameters [30]. The approach based on multifrequency NIM (m-NIM) exploits the advantages of synchronous impedance measurement at a given frequency range. Its practical implementation and the interpretation of results differ in comparison to DEIS.

In this work, the principal aspects of m-NIM methodology will be discussed. The object of our studies were heavy boron-doped diamond (BDD) electrodes. The BDD electrodes recently gained tremendous popularity in electrochemical studies due to their unique properties, such as wide potential window in aqueous electrolytes, high chemical stability to aggressive acids and bases, low background currents [31–35]. Due to their high sensitivity, the BDD electrodes are preferable with regard to electrochemical sensor applications [36–39].

Boron-doped diamond electrodes are grown using various CVD methods. The key aspects of the elctrodes, such as donor concentration, layer thickness, morphology and crystallite size, defects distribution, and the occurrence of *sp$^2$*-carbon contamination [40] are affected the most by the external conditions of CVD process. A heavy boron concentrations in gas admixture increases the electrode's conductivity, concurrently accelerating the formation of new nucleation sites and affecting the size of BDD crystallites [41]. Among many factors, which significantly affect the heterogeneous distribution of electric and electrochemical properties on the surface of BDD electrodes, are: (i) *sp$^2$*-carbon contamination from CVD process [42]; (ii) preferential boron accumulation at the grain boundary region [30]; (iii) boron dopant density [43]; (iv) crystallographic structure [44], and (v) the type of BDD electrode termination [45].

Surface termination plays a key role in the applicability of BDD electrodes. Hydrogen-terminated (HT-) electrodes are hydrophobic, non-polar and possess high electric conductivity [46–48] in contrast to hydrophilic oxygen-terminated (OT-) electrodes whose conductivity is also significantly decreased [46,48,49]. A large difference in conductivity results from the acceptor role of hydrogen atoms [21]. Also, the potential window is much wider for OT- electrodes compared to HT-BDD, shifting the energy bands to lower values [21,50,51]. As a result, HT-BDD are more prone to functionalization with various compounds for sensory purposes, which is particularly valuable in case of macromolecular compound detection [38–40,52,53].

The different surface terminations of BDDs show various electrochemical responses towards redox species. This enables the selective detection of specific target compounds [54]. Ivandini *et al.* [55] reported that oxalic acid could be determined only at hydrogen-terminated BDD surface. Watanabe *et al.* [56] presented that the electrochemical performance of HT-BDD deteriorates gradually when applied constantly without additional pretreatment. It was also noticed that the energy levels of the valence and conduction bands in HT-BDD are higher than in OT-BDD, causing the decrease of the thickness of the depletion layer in comparison to OT- BDD [57]. Next, Hamers and co-workers revealed that HT-BDD could be applied to generate solvated electrons by injection of UV light excited electrons in aqueous electrolyte [58]. Hydrogen termination of surface is crucial for efficient covalent grafting of biomolecules (e.g. DNA) [59], the immobilization of metallic nanoparticles [60] or direct amination for biomolecular interactions [61].

To complicate things further, Pleskov et al. [44,62] revealed that the oxidation of the termination bonds of epitaxially grown BDD layer on Si surfaces leads to the significant modification of surface electric properties in terms of donor density and flat-band potential. Crystallographic orientation is thus said to be the cause of higher charge transfer efficiency for Fe(CN)$_6^{3-/4-}$ redox species on (111) and (110)-oriented planes, while the charge transfer process is activation-controlled rather than diffusion-controlled on (100)-oriented planes [44]. Ryl et al. [63] demonstrated that the occurrence of similar behavior on polycrystalline BDD is the result of multistage oxidation process. The mechanism of the process is based on various propensity for oxidation among the certain types of crystal orientation, with the suggested order to be as follows: (110) > (100) > (111) [45]. Early AFM/NIM measurements revealed that polarization of +1.6 V vs Ag|AgCl is sufficient to oxidize (110)-oriented planes in 1M H$_2$SO$_4$, while other surface areas required much higher anodic polarization potentials.

There are many reported ways to affect the termination of BDD surface. The most commonly applied procedures are, among others, boiling in concentrated acidic solution [46,64], exposure to deep anodic polarization [36,63,65], exposure to oxygen plasma under UV radiation [48,66,67] as well as high-temperature [68,69] or ozone treatments [68]. Furthermore, the surface of BDD electrodes undergoes gradual changes as a result of natural ageing under UV radiation in the presence of atmospheric air [70,71]. In some reports, the utilization of oxygen plasma is said to be the most efficient, also leading to the modifications in the surface structure and morphology [52].

The most commonly used procedure for assessing the degree of surface modification is based on the contact angle measurements and/or surface chemistry analysis [72,73] According to Ghodbane et al. [74], the oxidation process results in termination with various functional groups, e.g. hydroxyl, carbonyl, carboxyl or ether in dependence on the type of oxidation treatment. Thus, a comparison of oxidation methods is problematic because the differences in surface composition, leading to various electrode hydrophilicity, have to be considered. The oxidation of BDD electrode dramatically increases its surface resistance, and charge transfer resistance in case of electrolytic environments [63,70,75]. Neither XPS nor the contact angle measurements give researchers the capability to investigate heterogeneity of surface oxidation on a microscopic scale. The local surface distribution of areas having various electric properties affects the electrochemical measurements through the modification of the diffusion field and might be the cause of reversibility issues [76]. Therefore, the goal of the presented research was to use m-NIM to measure the local changes in surface resistance, occurring as a result of the modification of BDD surface termination type and to compare different oxidation methods.

## 2. Material and methods

### 2.1 Highly boron-doped diamond electrodes

The BDD electrodes were synthesized in a microwave plasma-assisted chemical vapor deposition (MWPACVD) system (Seki Technotron AX5400S, Japan) on p-type silicon substrates with (100) orientation. The substrates were seeded by sonication in nanodiamond suspension (crystallite size of 5 – 10 nm) for 30 min. The detailed process parameters can be found elsewhere [21]. The samples were doped by using diborane ($B_2H_6$) at the [B]/[C] ratio of 10000 ppm in gas phase. The thickness of deposited film after a 6-h growth was approx. 3 μm [77].

### 2.2 Modification of BDD surface termination

The following pretreatment procedures were carried out on the investigated BDD samples:

*Hydrogen termination pretreatment.* First, each BDD electrode undergo pretreatment in high power hydrogen plasma. The MWPACVD system was utilized for that process. BDD surfaces were exposed to hydrogen plasma treatment for 10 mins. The flow of hydrogen was set to 300 sccm, temperature was 500ºC and the microwave power was 1300 W [78,79]. Next, the samples were subjected to one of the oxidation procedures, as described below.

*Electrochemical oxidation.* Anodic polarization exceeding +1.5 V is said to lead to the surface oxidation of BDD electrodes [50,80,81]. In the present study, the BDD samples were subjected to anodic polarization in 1M $H_2SO_4$ for a period of 30 min. Polarization potential was +2.5 V vs Ag|AgCl. The procedure was previously described in details in the literature [45], revealing varying propensity to modify the termination type in the BDD crystallites in dependence on their crystallographic orientations. The modification was performed using Autolab 302N potentiostat (Metrohm, The Netherlands).

*Chemical oxidation.* Samples were exposed to the mixture of concentrated $H_2SO_4$ and $KNO_3$ at the weight ratio of 2:1. Then, the solution was heated under the fume hood to 300 °C for one hour. The samples were then removed and cooled down in the air. Finally, they were boiled again in pure $H_2SO_4$ to remove soluble salts from the sample surface [49,82]. A similar procedure is often applied to decontaminate the electrodes after CVD [83].

*High-temperature oxidation in the air.* Electrodes placed in the furnace were baked at 600 ºC for 10 min and then cooled down in the air [49,84].

*Oxygen plasma treatment.* This approach is considered by some researchers to be the most efficient oxidation method so far [52,73]. For the purpose of this study, the electrodes were irradiated with a 50 W UV lamp for 10 min. The flow of ozone stream was 15 sccm. The process pressure was maintained at 0.6 mbar. [73].

*Ageing in atmospheric air*. The modification of BDD surface termination via ageing in atmospheric air occurs gradually over time, although changes in the electrochemical properties of BDD already after several weeks of exposure were described in some reports [70,71]. Nevertheless, it was reported that hydrogenated BDD electrodes display high electrochemical reactivity even after several months in storage [85,86]. Therefore, in this study, the samples were kept in a Petri dish under laboratory conditions, i.e. at room temperature and *in situ* humidity, for a period of 24 months.

**2.3 Setup for impedance measurements in nanoscale**

Local impedance studies were conducted by means of a commercial AFM device Ntegra Prima (NT-MDT, Russian Federation) supplemented with a setup for independent electrical measurements. The aforementioned setup consisted of PXIe-4464 and PXIe-6124 (National Instruments, USA) cards for AC signal generation and acquisition, respectively, operating in PXIe-1073 chassis. For the current measurements, the SRS 570 current preamplifier (Stanford Research System, USA) was used. To ensure good quality of the electrical contact, CDT-NCHR (Nanosensors, Switzerland) probes were used. Their geometrical and mechanical properties were as follows: probe dimensions 125×28×4.2 μm (L×W×T), tip radius 200-300 nm, resonance frequency 508 kHz, spring constant 130 N/m. The contact force determined on the basis of the landing curves was equal to 6.3 μN. The conductive layer on these probes is made of boron-doped diamond, which ensures high mechanical durability with moderately high conductivity [87,88].

The control software has been implemented in LabVIEW environment (National Instruments, USA), with a number of functional components. The first of them (generator) is designed to produce and continuously apply the voltage perturbation between the sample and the tip of the probe. The second component is the current response analyzer, responsible for the analysis of the subsequent fragments of recorded waveforms in real time. The third component was the impedance fitting software, based on Nelder-Mead algorithm [89], which has been necessitated by the large number of obtained spectra in the scanning mode. The final component was used for estimation of the investigated surface area coverage owing electric parameter above given threshold.

**2.4 Physicochemical examination of surface termination**

*X-Ray Photoelectron Spectroscopy (XPS)* analysis. High-resolution XPS spectra were recorded in the *C1s* and *O1s* binding energy (BE) range for each investigated sample to verify the changes in surface chemistry for various types of oxidation. The measurements were carried out using Escalab 250Xi multispectroscope (ThermoFisher Scientific, United Kingdom), equipped with Al Kα monochromatic X-Ray source, spot diameter 650 μm. Applied pass energy was 10 eV, and the energy step size was 0.1 eV. Charge compensation was controlled through low-energy electron and low-energy Ar$^+$ ions emission using the flood gun (emission current 150 µA, beam voltage 2.1 V, filament current 3.5 A) [91]. The peak deconvolution analysis was performed using Avantage software provided by the manufacturer.

*Contact angle analysis.* Measurements were performed on a Drop Shape Analyzer DSA100 (Krűss, Germany). The size of a water drop was 5 µL, while the contact angles were automatically measured by ADVANCE software provided by the manufacturer. The measurement was repeated three times per sample.

## 3. Multifrequency Nanoscale Impedance Microscopy (m-NIM) methodology specification
### 3.1. Perturbation and response signal

The scheme of m-NIM measurement setup is presented on **Fig. 1a**. Multifrequency voltage perturbation signal is applied between the tip of the probe and investigated material. The signal is composed of a superimposed package of elementary perturbation waveforms, used simultaneously as shown in **Fig. 1b**.

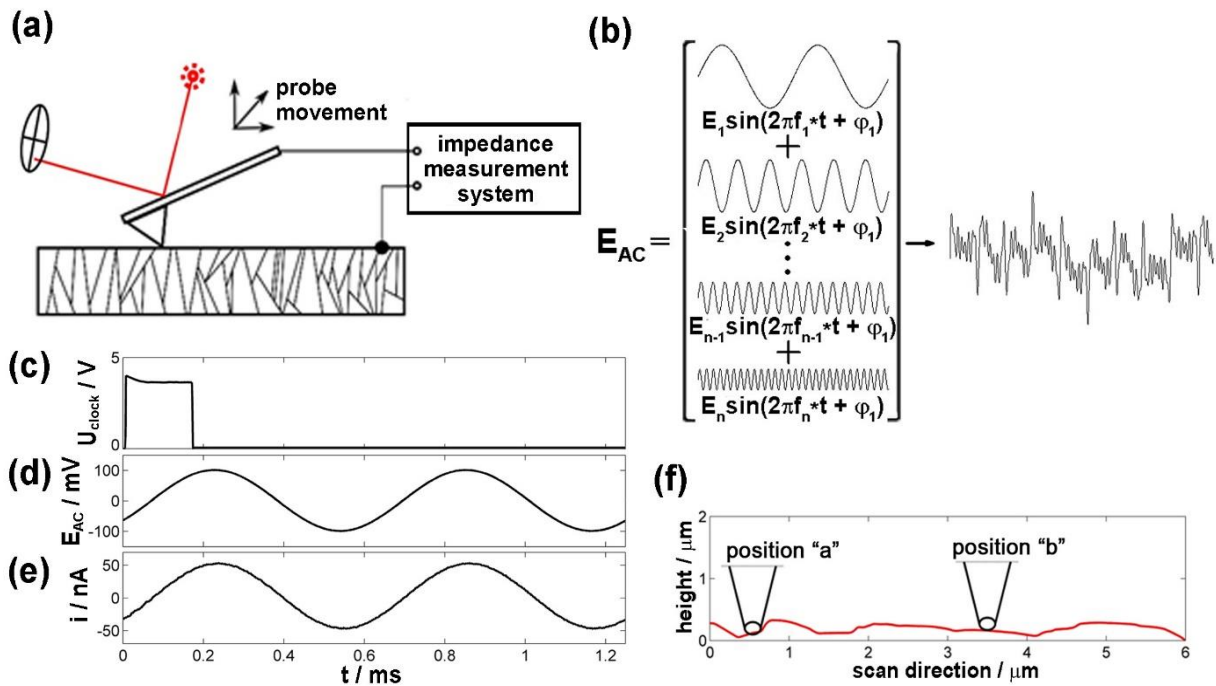

**Fig. 1** – (a) schematic representation of Multifrequency Nanoscale Impedance Microscopy measurement setup, (b) composition of multifrequency voltage perturbation signal by using the pack of elementary signals, (c-e) idea of synchronization between perturbation E$_{AC}$, response i and probe localization by means of clock impulses U$_{clock}$ generated by AFM controller at the beginning of height acquisition at each pixel, (f) scheme of tip tapping on the BDD surface, reflecting the scale of contact area and its change during a scan.

The amplitude $E_i$, frequency $\omega_i$ and phase-shift $\tau_i$ of each elementary waveform were selected individually. The distribution of frequencies was obtained by introducing prime numbers as multiples of the fundamental frequency. The above-mentioned operation was aimed at avoiding the overlapping of measured frequencies and overtones generated by possible non-linearity of the current-voltage characteristics [22]. The upper applied frequency limit must be much lower than half the sampling frequency, while the lower frequency is related to total data acquisition time, which in this case translates to AFM scan rate. A global modification of the effective signal amplitude through shifting phases of individual components was performed to maintain it below 100 mV. This procedure was carried out using an iterator program written in LabView environment. An additional DC bias voltage 0.5 V was applied to multifrequency perturbation signal in order to decrease the resistance of semiconductive probes.

The general concepts of the waveform registration process for the impedance determination are presented in **Fig. 1c-e** One of analog-to-digital converter measuring channels was used to trigger the measurement by means of a clock signal generated by the microscope controller before starting the height measurement for each pixel of the image (**Fig. 1c**). Simultaneously recorded perturbation voltage value (**Fig. 1d**) and current response (**Fig. 1e**) in the form of finite data sequences are the subject of further spectral analysis. The illustration only shows single frequency signal component to emphasize how the perturbation frequencies have been selected.

The registered signal is sampled with the defined sampling frequency $f_s$ and subsequently quantized using an analog-to-digital converter with defined resolution (here: 24 bit). The resultant signal is then cut using Short-Time Discrete Fourier Transformation (STDFT) to acquire local impedance spectra at each analyzed pixel. Through the use of multifrequency perturbation, it is possible to record the impedance spectrum in the selected frequency range for each pixel of the topographic image. The analysis is performed in real time, therefore it is possible to observe the impedance image and verify its correctness. In the imaging mode, the registration of current and voltage signals is performed synchronously with the acquisition of topographic image, hence the correlation of the topographic and impedance maps is provided.

An important issue arises due to the geometry of probe/sample contact and real interaction surface between these two, as schematically presented in **Fig. 1f** [92]. The changes in impedance may be a result of both the heterogeneity in electric parameters and variable contact surface between the probe tip and rough surface. When the probe tip is located in position "a", the value of surface resistance is affected by enhanced contact area between the side walls of the probe and the effects of topography on the sample surface. Thus, it is important for discussion purposes to estimate the effect of additional geometric factor on the measured impedance value, taking into consideration the average geometric dimensions of the investigated object. In the situation illustrated as location "b" in **Fig. 1f**, the value of contact surface was estimated taking into account the surface indentation in the form of a spherical part of the tip. The Johnson-Kendall-Roberts mechanical contact model was used to calculate the force causing the indentation as 6.3 μN. Based on the literature and measured data ($E_{tip}$ = 179 GPa, $E_{samp}$ = 1 GPa, $\upsilon_{tip} = \upsilon_{samp}$ = 0.3 μm, $r_{tip}$ = 200 nm, $r_{samp} \approx \infty$), the contact surface was estimated to be 0.059 μm$^2$, [93]. On the other hand, in the case illustrated as location "a", the contact surface resulting from the contact of the pyramidal side of the tip with rough sample surface was 0.022 μm$^2$. The following values of parameters were used in calculations: $r_{tip}$ = 200 nm, tip aspect ratio 5:1, the mean roughness of BDD sample equal to h = 50 nm.

The difference between the two contact surface areas described above is within one order of magnitude. Therefore, the changes in impedance reaching several orders of magnitude, as reported in this study or in previously published literature [94], cannot be solely explained by the changes in surface geometry.

### 3.2. Impedance spectra analysis and m-NIM map generation

In the paper two possibilities of impedance registration is presented namely with stationary position of the probe (referred as spectroscopic mode) and during topographic scanning (imaging mode). Thus, a single spectrum or a set of spectra is obtained as a result of the impedance measurement. Regardless of the measurement regime, it is necessary to fit the obtained spectra using previously assumed electric model (electric equivalent circuit, EEC). The registration of multifrequency spectrum for each pixel leads to the creation of large amounts of data (256 by 256 pixels or $2^{16}$ impedance spectra). The fitting procedure was carried out using dedicated component implemented to measurement software and based on Nelder-Mead algorithm, thus making it possible to obtain maps of the distribution of the EEC parameters on the analyzed BDD surface.

In our study, we used the $R_S(CR_P)$ EEC suggested in the literature (see inset of **Fig. 2**). The EEC consisted of three elements, i.e. serial resistance ($R_S$) representing a sum of spreading resistance of sample and probe as well as resistance of the tip conductive layer; the tip/sample contact resistance ($R_P$); and contact capacitance (C). In the hardware configuration applied, the determined capacitance also included the parasitic components originating from the side surfaces of the AFM probe and the cantilever [90]. Thus, it was not possible to determine the absolute value of capacitance. The local changes of capacitance depend on the surface distribution of material's properties.

It should be noted that the largest changes were observed in the values of parameter $R_P$, associated with contact properties between the tip of the probe and the sample surface. This parameter is primarily associated with the changes in the sample's surface resistance, being the only variable. The measured changes concern the capacitance and serial resistance to a lesser extent. The accuracy of small capacitance values measurement is limited, thus the separation of individual resistance components during the fitting procedure can also be subject to uncertainty. Thus, the total resistance $R_{S+P}$ value seems to be appropriate for assessing the relative conductivity changes of BDD. Its change is primarily affected by modification of surface resistance under various oxidation treatments.

When applied to the scanning AFM mode (impedance imaging), the spectral analysis takes place in real time, operating on data recorded synchronously with the topography acquisition of each pixel in an image. This necessitates the adjustment of the length of the data sequence, and thus the duration of the basic period of multifrequency perturbation, to the topographic scan rate. The above assumption enforces a significant limitation in the low frequency range when measuring in the scanning mode.

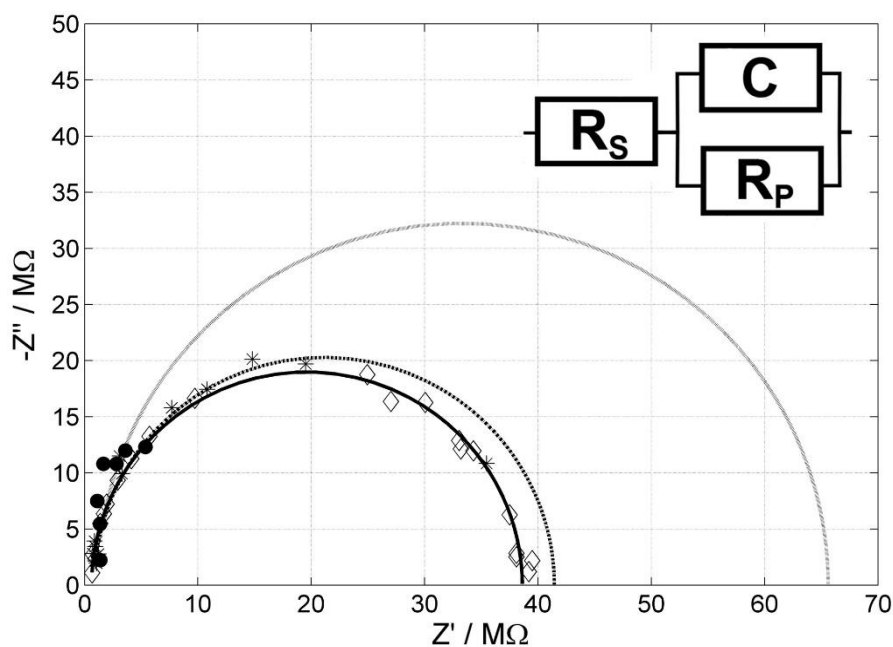

**Fig. 2** – Comparison of spectra obtained in various conditions: (◇) probe in fixed position, and imaging mode with topographic scan rate 0.1 (✳) and 1 (●) Hz. OT-BDD sample after electrochemical oxidation.

A comparison of exemplary impedance spectra, resulting from both applied approaches, is presented in **Fig. 2** and corresponding electric parameter values are summarized in **Table 1**. These results were performed or OT-BDD electrode after electrochemical treatment. It is clearly visible that during scanning, a significant amount of impedance data within low frequency range is irreversibly lost due to the measurement procedure demands. The limited amount of data suggests caution in the context of EEC fitting procedure, while much higher fitting errors are expected due to the uncertain location of low frequency data intersection.

**Table 1** – The calculated values of electric parameters on the base of $R_S(CR_P)$ EEC analysis of impedance spectra presented on **Fig. 2**. OT-BDD sample after electrochemical oxidation.

|  | $R_S$ / MΩ | C / pF | $R_P$ / MΩ | $R_{S+P}$ / MΩ | $\chi^2 \times 10^3$ |
|---|---|---|---|---|---|
| fixed position | 0.61 | 13.79 | 38.06 | 38.67 | 4.9 |
| imaging (scan rate 0.1 Hz) | 0.84 | 7.83 | 35.50 | 36.34 | 18.0 |
| imaging (scan rate 1 Hz) | 1.11 | 4.15 | 64.09 | 65.20 | 98.9 |

In order to extend the frequency band of the impedance measurement, attempts were made to decrease the topographic scan rate. This allowed us to obtain a better fit of the impedance spectra, while the downside was the deterioration of the surface image quality, see **Fig. 3**. It should be noted that the aforementioned effect was not compensated by means of horizontal position sensors embedded in the piezoelectric scanner, which indicates a hardware limitation of the device used.

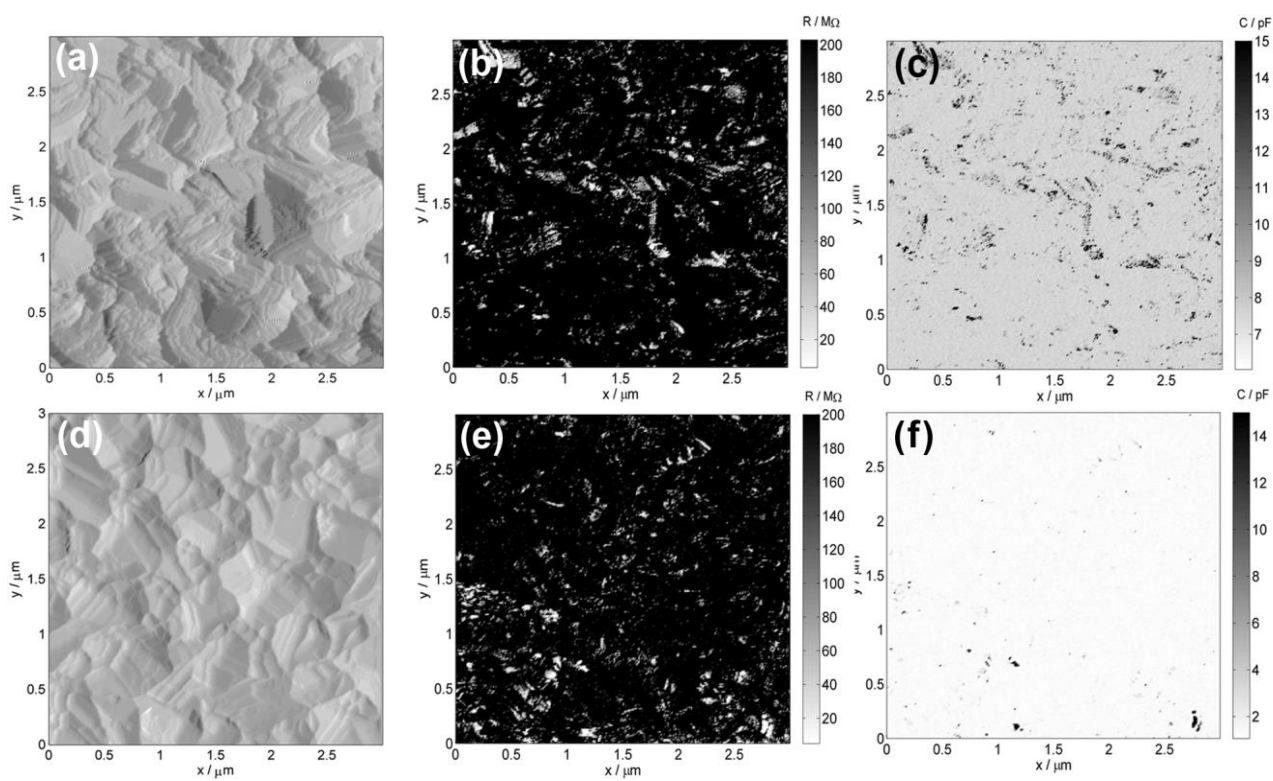

**Fig. 3** – AFM and m-NIM results using two different scan rates (a-c) 0.1 Hz and (d-f) 1 Hz, (a,d) AFM contour micrograph in DFL mode, (b,e) distribution of $R_{P+S}$, (c,f) distribution of capacitance C. OT-BDD sample after electrochemical oxidation.

The pixel-wise change of contact point introduces significant distortion to the recorded current value, which can be observed as increased spread of impedance data, influencing the determined electric parameter values as a consequence (see **Fig. 3e,f** for comparison).

The observed variation of capacitance parameter is much smaller for BDD material as it is affected the most by changing the scan rate of AFM probe and changes in the contact surface area. Furthermore there is only slight difference between the measured capacitance of HT-BDD and OT-BDD, ranging from ones to tens pF. Thus, in the next section of this paper, only the surface resistance maps will be presented.

## 4. Results and discussion
### 4.1 m-NIM surface resistance maps for various types of BDD surface modification

*Hydrogen-terminated BDD electrode*. The AFM contour image and m-NIM map are presented in **Fig. 4**. In the previous studies, we reported that the surface resistance of HT-BDD electrodes does not exceed several M$\Omega$ (typically, in k$\Omega$ range) [17,45,94–96]. The grain boundary regions of BDD are characterized by much higher boron concentration and thus even higher conductivity [17,94]. It is clear that the vast majority of examined HT-BDD sample surface area falls below 10 M$\Omega$.

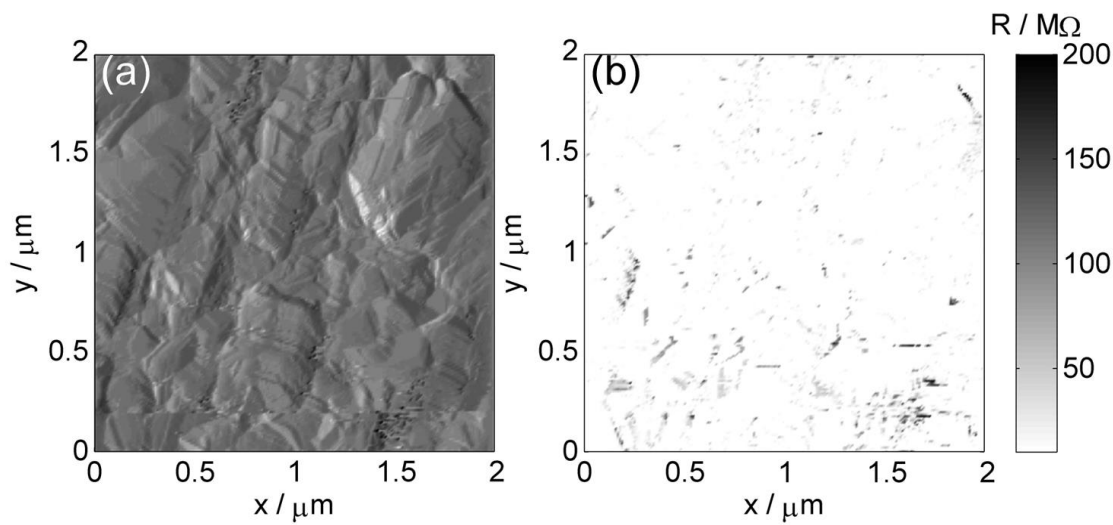

**Fig. 4** – (a) AFM contour micrograph in DFL mode, and (b) total $R_{S+P}$ resistance map for HT–BDD. Reference sample pretreated in hydrogen plasma.

On the other hand, the local resistance values increase as a result of surface oxidation and can vary between several dozen M$\Omega$ up to even G$\Omega$, where measuring equipment sensitivity hinders its proper evaluation. Thus, in order to determine the electrode modification rate with each investigated approach we propose to calculate the percentage of surface coverage for areas having the total resistance falling below 10 M$\Omega$ set as a threshold for non-modified hydrogen termination. In the case shown on Fig. 3 only around 4.4 % of the analyzed HT-BDD surface area is characterized by $R_{S+P} > 10$ M$\Omega$.

*Electrochemical oxidation*. The AFM contour micrograph as well as m-NIM surface resistance map of BDD sample after electrochemical oxidation in 1M $H_2SO_4$ at +2.5 V vs Ag|AgCl are shown in

**Fig. 5**. The investigated procedure was a subject of previous studies by Ryl et al. [45], who concluded that the oxidation propensity of particular BDD crystallites differs, and that it depends on anodic polarization depth as well as crystallographic orientation. It is important to note that the surface oxidation of (110)-oriented planes is initiated at much lower polarization potentials, exceeding +1.5 V under these conditions. The micrograph in **Fig. 5b** reveals the structure which is similar to the one reported previously, evidencing subtle local changes in surface resistance due to heterogeneous modification of termination type. White color on the m-NIM map is associated with the areas where the estimated surface resistance value is below the threshold of 10 MΩ. In any other region of the map, the surface modification based on the transition from hydrogen termination to oxygen termination took place. It is clearly visible that the measured resistance has significantly increased as a result of the electrochemical treatment. The location of oxidized areas strongly coincides with the shapes of BDD crystallites. According to previously presented studies, (111)-oriented planes were recognized as having the weakest propensity for oxidation. It should also be noted that the electrode oxidation via anodic polarization does not lead to a uniform surface modification under the aforementioned conditions. About 11 % of the analyzed BDD surface remained hydrogenated.

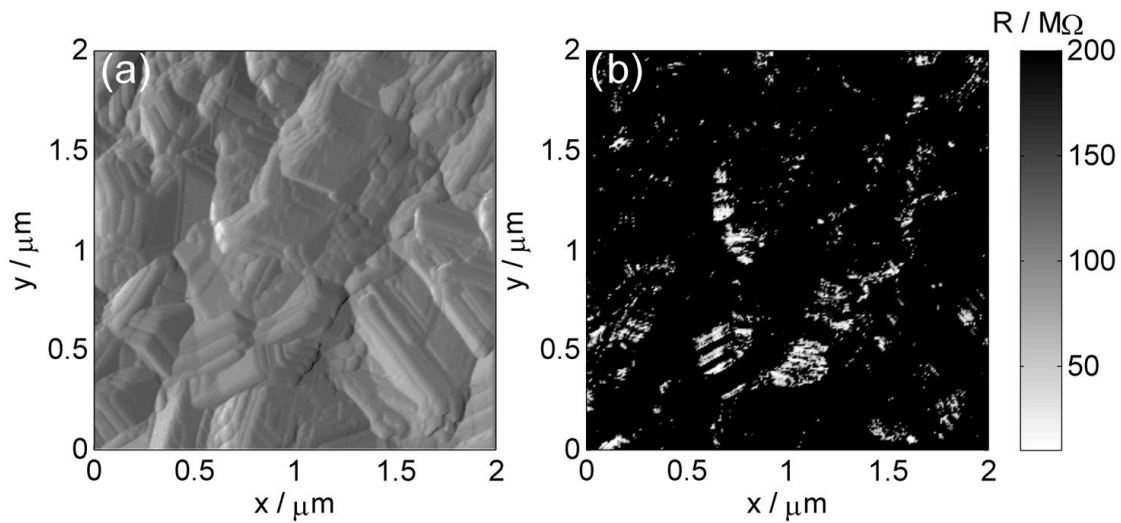

**Fig. 5** – (a) AFM contour micrograph in DFL mode, and (b) total $R_{S+P}$ resistance map for OT–BDD sample exposed to anodic polarization in $H_2SO_4$ at +2.5 V vs Ag|AgCl.

*Chemical oxidation*. The analysis of the BDD electrode after the chemical treatment in boiling concentrated $H_2SO_4$ + $KNO_3$ solution revealed similar result of surface oxidation yet much more uniform modification, as can be seen in **Fig. 6**. The estimated share of hydrogenated surface area was considerably smaller, i.e. not exceeding 1.5 % of the analyzed surface area. It is clear, however, that the small area of hydrogenated surface visible in the bottom right corner of **Fig. 6b** resembles the shape of BDD crystallite.

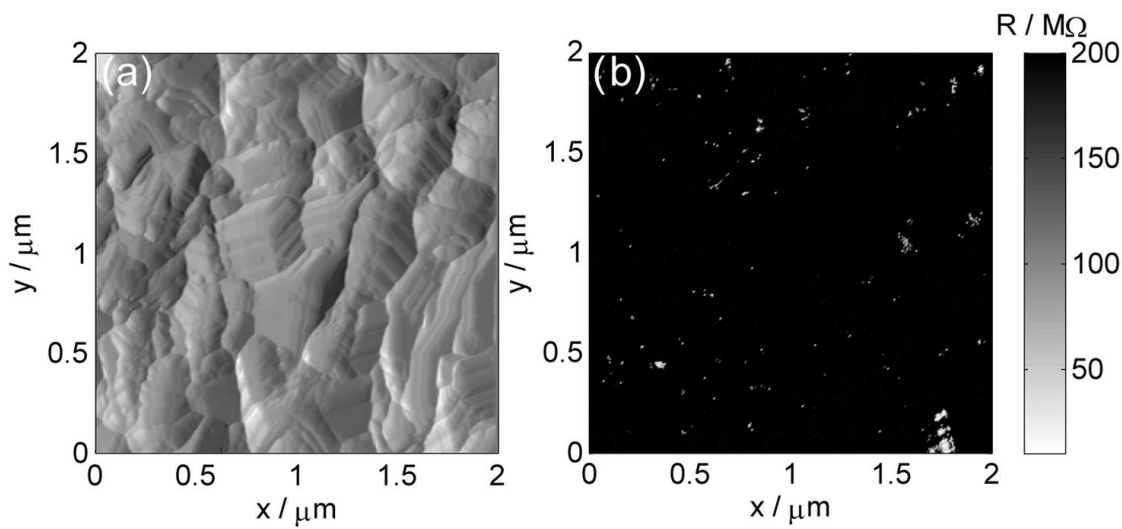

**Fig. 6** – (a) AFM contour micrograph in DFL mode, and (b) total $R_{S+P}$ resistance map for OT–BDD sample exposed to boiling at 300 °C in $H_2SO_4$ + $KNO_3$ for 1 hr.

*High-temperature oxidation*. Contrary to previously presented BDD oxidation methods, the sample exposure to high temperature (600 °C) in electrolyte-free environment resulted in the most heterogeneous surface in terms of the distribution of electrical properties. The effect of the treatment is clearly visible on the resistance maps obtained by means of m-NIM (see **Fig. 7b**). The share of modified electrode surface area barely reached 56%. As in the case of previously presented results, the surface resistance values strongly vary among particular BDD crystallites. It should be concluded that the electrode modification due to exposure to high temperature is very selective with regard to plane orientation. It is highly likely that the crystals oriented in one or more major planes did not oxidize under the above conditions, since the texture of investigated BDD electrodes was composed primarily of (110) and (111) facets.

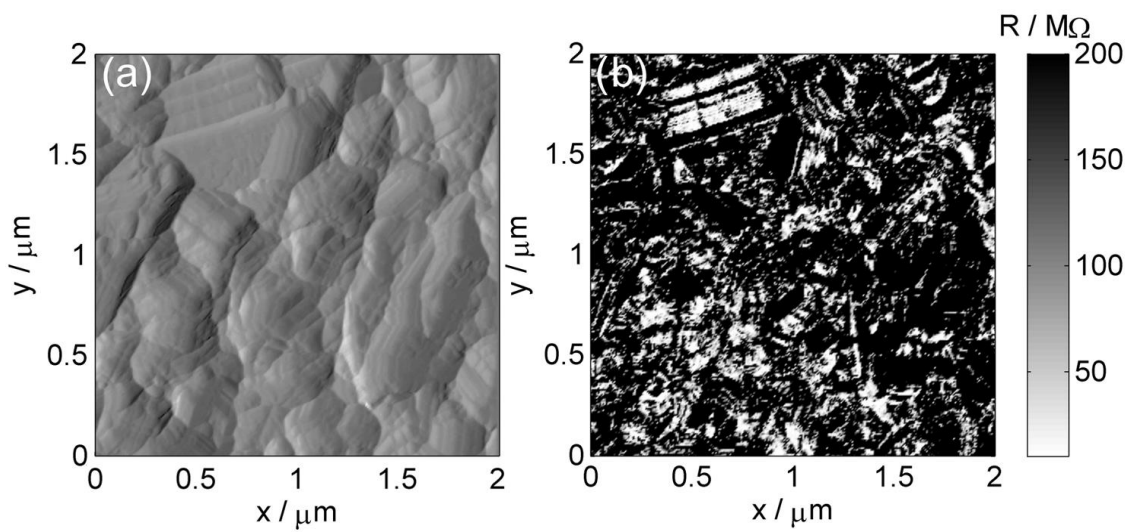

**Fig. 7** – (a) AFM contour micrograph in DFL mode, and (b) total $R_{S+P}$ resistance map for OT–BDD sample exposed to heat in furnace at 600 °C for 10 min.

Oxidation at elevated temperatures proceeds in two phases and it accelerates the abstraction of surface hydrogen. This, in turn, accelerates the adsorption and desorption of oxygen as carbonyl or carboxyl and, in some cases, almost a complete removal of hydrogen termination. These modified surfaces may be dominated by defects with very different chemistry compared to untreated surfaces [97].

*Oxygen plasma treatment.* Based on m-NIM maps, the degree of BDD surface modification due to treatment was very high and quite homogeneous (see **Fig. 8**). The surface modification level of 97.7% places the oxygen plasma treatment above other oxidation methods used, chemical oxidation being the only comparable treatment (98.5 %). Nevertheless, the presence of areas considered to be non-modified coincides with the topography of individual BDD crystallites, which was also observed for other treatments. In the case of oxygen plasma method, the efficiency of surface termination modification was primarily affected by the length of oxidation [73]. m-NIM studies corroborate the previous reports revealing that the oxygen plasma treatment is more efficient than the electrochemical approach [52,98].

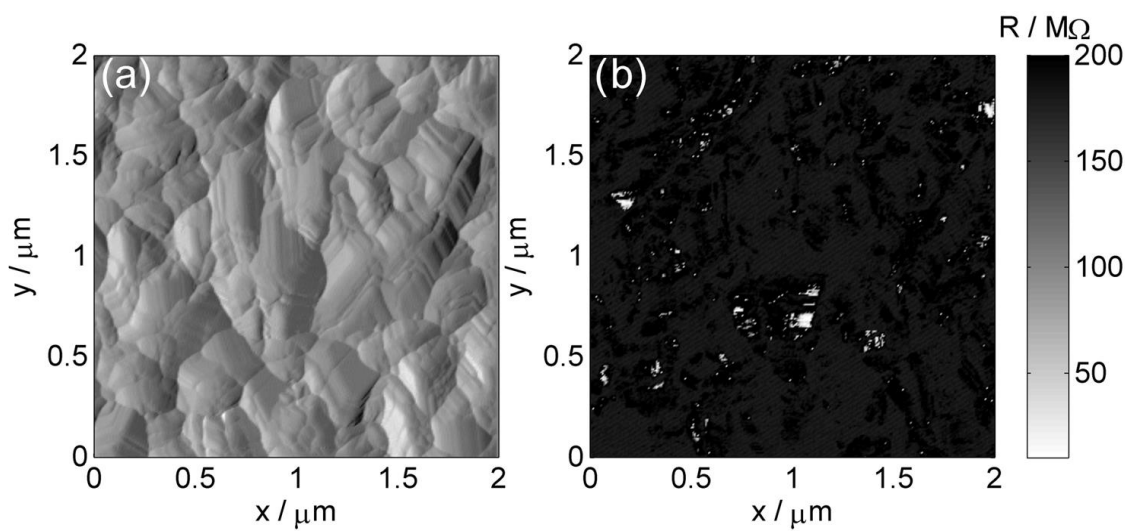

**Fig. 8** – (a) AFM contour micrograph in DFL mode, and (b) total $R_{S+P}$ resistance map for OT–BDD sample exposed to oxygen plasma and 50 W UV lamp for 10 min.

*Ageing in atmospheric air.* We have discovered that regardless of the reported high chemical stability, a prolonged exposure of BDD electrodes to atmospheric air significantly alters their electric properties. **Fig. 9b** shows a prominent increase in surface resistance as a result of such a treatment, with the average value of surface resistance significantly exceeding the applied threshold that identifies modification to OT-BDD. Given 10 MΩ as soft oxidation marker, the electrode modification was quite deep (81.6% surface coverage), while the distribution of non-modified surface regions was quite homogeneous (**Fig. 9a**). Geisler and Hugel have also observed that ageing affects the surface termination of BDD, and concluded that it originates from physisorption of atmospheric adsorbates and long-term conversion of surface chemical groups [71]. This type of oxidation is also considered to decrease electrochemical reactivity [70].

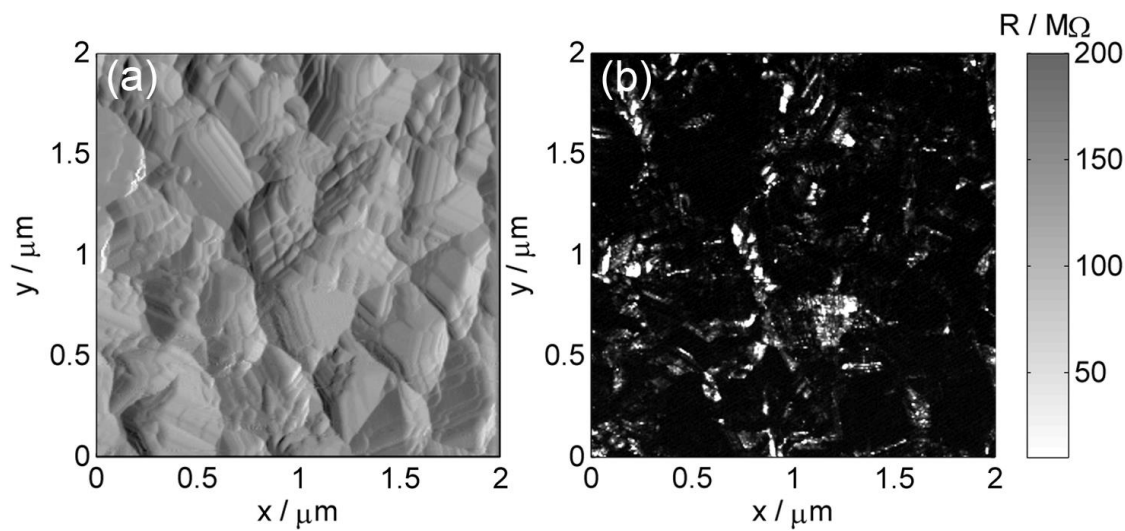

**Fig. 9** – (a) AFM contour micrograph in DFL mode, and (b) total $R_{S+P}$ resistance map for BDD electrode aged in atmospheric air at 25 °C for 24 months.

Diverse activation energies are needed for graphitization and surface modification due to the various structures of the (100) and (111) diamond facets [99]. The structure and stability of diamond modifications were also studied by Petrini and Larsson [100,101], who reported that the adsorption energy values of H, O, and OH on the diamond (100) facets are 4.53, 5.28, and 4.15 eV, respectively. In contrast, the adsorption energy values of H, O, and OH on the diamond (111) facets are 4.13, 6.21 and 4.30 eV, respectively. Thus, the adsorption of oxygen and the carbonyl formation are energetically favored at the (100) facets. On the other hand, the molecular dynamics simulations indicate that all three low-index surfaces can be etched by hyperthermal atomic oxygen. The etching yield of diamond surfaces indicates that the (110) facet is the least resistant and the (100) facet is the most resistant surface, while etching proceeded by simultaneous functionalization of the surface with oxygen atoms [102].

**4.2 Examination of the degree of surface modification**
*High-resolution XPS analysis.* Samples were analyzed in the Binding Energy (BE) range of *C1s* spectra in order to distinguish the chemistry of BDD surface termination resulting from various oxidation treatments. The proper interpretation of XPS spectra recorded for polycrystalline BDD electrodes was difficult due to the high density of intergranular defects and grain boundaries. The position of the primary component in the analyzed spectra may shift by up to 1.0 eV with increasing [B] dopant density (from ~$10^{20}$ to $10^{21}$ cm$^{-3}$), which mainly results from the presence of $sp^2$-C at the grain boundaries and in structural defects [103]. Furthermore, it was previously reported that crystallographic orientation is also extremely important [103,104].

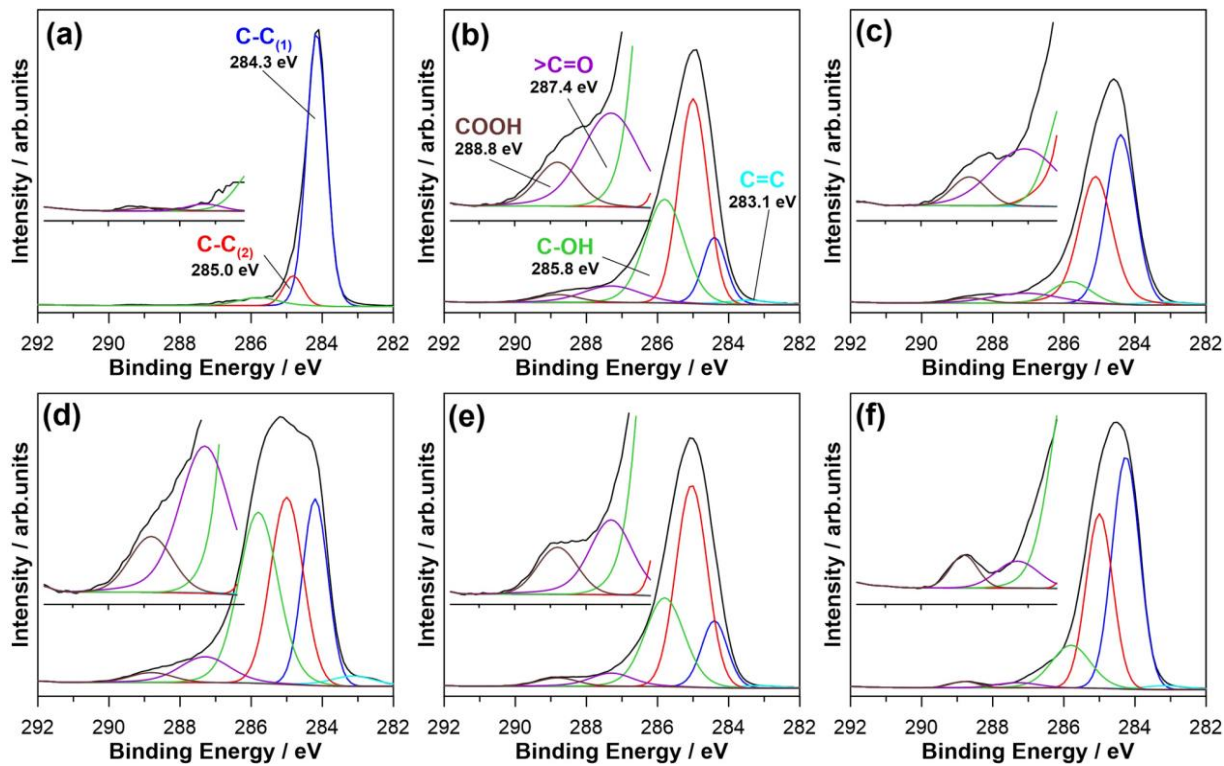

**Fig. 10** – High-resolution XPS spectra in the *C1s* binding energy range recorded for (a) HT-BDD, and (b-f) OT-BDD electrodes after various oxidation treatments, i.e. (b) electrochemical, (c) chemical, (d) high-temperature, (e) oxygen plasma and (f) prolonged air exposure. Magnified spectra fragments (x8) in the inset.

Each spectrum has been deconvoluted to identify the share of various components, as shown in **Fig. 10**. Given that the investigated electrodes fall in the range of heavy doping (2 x $10^{21}$ cm$^{-3}$) and the (110) and (111) facets dominate their texture [17,105–107], the primary component on HT-BDD surface was located at 284.3 ± 0.1 eV. The C-C$_{(1)}$ component should be attributed to $sp^3$-C CH species on the hydrogenated diamond surface as well as to $sp^3$-C from the 'bulk' [63,108,109]. The second most noticeable component, i.e. C-C$_{(2)}$ is shifted by +0.7 eV. It corresponds to polyhydride carbon species (CH$_x$) adsorbed onto $sp^3$-C surface and non-hydrogenated BDD surface, whose contribution significantly increases for any type of OT-BDD surface [63,110]. Three different types of oxygenated groups at the surface of investigated electrodes, namely, hydroxyl, carbonyl and carboxyl species were all deconvoluted from *C1s* spectrum using peak singlets. The values of peak BE for these species were found, i.e. 285.8 eV for C-OH and C-O-C, 287.4 eV for >C=O, and 288.8 eV for COOH, which were in good agreement with the available literature data for highly boron-doped electrodes [39,63,67,92,103,110]. Finally, a small $sp^2$-C contribution was found at low BE, shifted by -1.2 eV in relation to the primary component. Its presence was most prominent for non-oxidized samples due to post-CVD process contamination and resulting from high-temperature oxidation (2.7%), which is also the most damaging. The results of surface chemical analysis are presented in **Table 2**.

Table 2 – Chemical composition of the surface of BDD electrodes based on high-resolution XPS analysis in the $C1s$ BE range.

| Type of modification applied to BDD surface | $C1s$ chemical states and respective BE, in eV | | | | | |
|---|---|---|---|---|---|---|
| | -C-C(1) 284.3 | -C-C(2) 285.0 | -C-OH 285.8 | >C=O 287.3 | -COOH 288.8 | -C=C 283.3 |
| hydrogen terminated | 83.4 | 8.7 | 4.8 | 0.6 | 0.4 | 2.1 |
| electrochemical oxidation | 12.1 | 44.6 | 31.5 | 7.5 | 2.5 | 1.8 |
| chemical oxidation | 43.4 | 40.4 | 7.4 | 5.9 | 1.5 | 1.4 |
| high-temperature oxidation | 23.5 | 30.5 | 34.4 | 6.8 | 2.1 | 2.7 |
| oxygen plasma treatment | 13.8 | 49.5 | 28.2 | 4.9 | 2.6 | 1.0 |
| prolonged air exposure | 38.5 | 48.0 | 8.9 | 2.3 | 1.5 | 0.8 |

HT-BDD sample had the smallest content of oxidized forms of carbon, resulting primarily from exposure to atmospheric air and the adsorption of adventitious carbon onto the electrode surface. The highest modification efficiency of surface termination type was observed in the case of electrochemical oxidation in $H_2SO_4$ and the plasma treatment under UV radiation (see Fig. 9b,e). The surface chemistry associated with these two oxidation types also appears to be similar, with the dominant influence of hydroxyl species (~30%) and the highest amount of carboxyl species (>2.5%) among all investigated samples. Anodic polarization also resulted in the high amount of carbonyl species (7.5%). A similar amount and ratio of oxidized species were also observed on the BDD surface after high-temperature treatment, while the main distinguishing characteristic was the high C-C$_{(1)}$ contribution (24%; two times higher than for the plasma and anodic treatments). Such a result is in perfect agreement with the m-NIM observations, testifying to more diversified propensity of particular BDD crystallites towards oxidation at high temperatures. The mechanism of chemical oxidation in $H_2SO_4$ + $KNO_3$ is completely different. It results in a relatively high share of non-hydrogenated carbon, yet the hydroxyl species are almost lacking in comparison to other oxidation types (7% vs ~30%). It is worth noting that carbonyl groups can be formed specifically on the (100)-facets and the edges of (111)-facets due to the interaction of the dangling bonds of surface $sp^3$-carbon atoms with •OH or via the graphitized surface [80]. Importantly, reported chemical states on the surface of oxidized BDD electrodes corroborate previous studies [111,112].

We also took the opportunity to investigate the stability of highly boron-doped diamond electrodes that had been aged under atmospheric conditions. It was found that the hydrogenated carbon to non-hydrogenated carbon ratio of the sample after a 24-month exposure significantly differed from that of HT-BDD sample. The most important finding was the detection of spontaneous electrode oxidation due to the aging process, while the sample chemistry resembled that of the surface after chemical oxidation, being primarily contaminated with carboxyl and, to a lesser extent, with hydroxyl species, and almost lacking carbonyl groups. The obtained XPS results corroborate the m-NIM data for the sample exposed to the atmospheric air.

*Contact angle measurements.* The surface wettability of BDD strongly depends on the functional groups that terminate its surface. The results of contact angle measurements are presented on **Fig. 11**.

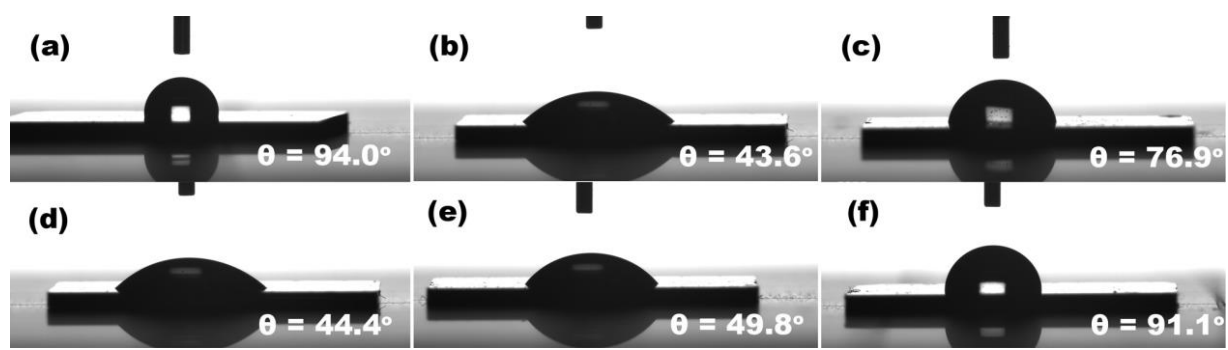

**Fig. 11** – Contact angle measurements for each investigated BDD electrode, i.e. (a) HT – BDD, and (b-f) after various oxidation treatments: (b) electrochemical oxidation in $H_2SO_4$, (c) chemical oxidation in $H_2SO_4$ + $KNO_3$, (d) high-temperature oxidation at 600 °C, (e) oxygen plasma treatment, and (f) long-time exposure to atmospheric air. Droplet volume was 5 μL.

The XPS analysis confirmed that the electrochemical anodization or plasma treatment of HT-BDD surface generates primarily hydroxyl (OH) groups, although some carbonyl and carboxyl groups also form. The content of carbonyl groups in the surface composition is much higher in case of thermal and chemical oxidation. The lowest values of contact angle were measured for electrochemical, high-temperature and plasma-based oxidation, reaching 43.6, 49.8 and 44.4, respectively. In case of chemical oxidation and air-exposed sample, the higher contact angle is attributed to the lower amount of hydroxyl groups on diamond surface. The most evident difference in the contact angle values among the analyzed BDD samples is due to the various polarity of bonds in the functional groups, i.e. hydroxyl, carboxyl and others [113–115].

The results on the degree of BDD termination type modification, assessed by various methods, are summarized in **Table 3**. The m-NIM approach towards the modification degree assessment was performed by calculating the surface area with $R_{S+P}$ resistance above the given threshold of 10 MΩ.

The XPS analysis based on the calculation of the total share of *C1s* peak components related to OT termination (namely, C-C$_{(2)}$, C-OH, >C=O and COOH; see **Table 2**) versus the components related to HT termination (namely: C-C$_{(1)}$ and C=C). These data were cross-verified with the outcome of contact angle studies.

**Table 3** – Comparison of the oxidation degree of differently terminated BDD based on m-NIM, XPS and contact angle analyses.

|  | m-NIM oxidized area / % | XPS analysis (C$_{OT}$:C$_{HT}$ share) | contact angle / deg. |
| --- | --- | --- | --- |
| hydrogen terminated | 4.4 | 0.2 | 94.0 |
| electrochemical oxidation | 89.0 | 6.2 | 43.6 |
| chemical oxidation | 98.5 | 1.2 | 76.9 |
| high-temperature oxidation | 56.4 | 2.8 | 44.4 |
| oxygen plasma treatment | 97.7 | 5.7 | 49.8 |
| prolonged air exposure | 81.6 | 1.5 | 91.1 |

It can be seen that the results obtained with each aforementioned approach are in agreement. The highest degree of OT-BDD surface coverage resulted from the electrochemical and oxygen plasma treatments. The contact angle measurement is the least accurate (yet the easiest to perform), which was proved by the analysis of a sample exposed to high temperature, showing a high degree of oxidation heterogeneity.

Furthermore, a decrease in the contact angle values was small for the chemically oxidized and air-exposed samples because the surface chemistry of these samples was lacking the hydroxyl groups that have low influence on the contact angle. The XPS and m-NIM studies effectively demonstrated that HT-BDD became modified to OT-BDD.

**Conclusions:**

We have described the methodology for performing the m-NIM analysis by means of AFM. The proposed m-NIM approach offers precise analysis of subtle changes in local electric properties on a submicroscopic scale. The primary advantage of our approach is the ability to obtain extended information about the electrical nature of the surface. In the case of impedance measurements, they contain both resistive and capacitive contributions which can be elucidated by means of fitting the impedance spectrum with EEC. The restrictions of the presented approach concern a much higher level of distortion in case of the scanning mode of measurement. The limited impedance acquisition rate in the low frequency range affects the capacitance measurements in particular, where the effective differentiation of ~ pF values was not possible during fast m-NIM scans. Moreover, the acquired impedance response is also affected by the process of scanning and consequently, by an abrupted contact between the probe and the sample.

Based on the presented comparison of various common procedures used to oxidize the surface of BDD electrodes, we can definitely state that the resulting modifications displayed different levels of heterogeneity. The oxidation degree was estimated as the increase in $R_{P+S}$, related primarily with surface resistance. We revealed that the oxidation propensity of polycrystalline electrodes coincides with the grain structure of BDD and thus it is most likely dependent on its crystallographic texture.

The XPS and contact angle measurements confirmed that various oxidation procedures lead to different surface chemistry of BDD which makes it more difficult to quantify the rate of termination type modification.

From among the investigated methods, m-NIM was the only approach that enabled the determination of local oxidation homogeneity, which is of great importance to electrochemical studies. Local conducting and insulating areas on the electrode surface will affect the charge transfer process and the diffusion field in electrolytic environment, therefore, the degree of diffusion field modification should be a topic of separate studies.

**Acknowledgments:**

The authors gratefully acknowledge financial support from the National Science Centre Grant Sonata No. 2015/17/ST5/02571 and National Centre for Science and Development Grant Techmatstrateg No. 347324 2015/16/T/ST7/00469. The DS funds of the Faculty of Electronics, Telecommunications and Informatics of the Gdansk University of Technology are also acknowledged.